\documentclass[a4paper,10pt]{article}
\usepackage[T1]{fontenc}
\usepackage[latin1]{inputenc}
\usepackage[nooneline,bf]{caption} 
\usepackage{times}
\usepackage{multicol}
\usepackage{amsmath}
\usepackage{amssymb}
\usepackage[dvips]{graphicx}
\usepackage{epic}
\usepackage{eepic}
\usepackage{epsfig}
\usepackage{xspace}
\usepackage{color}
\usepackage[subrefformat=simple, labelformat=simple]{subcaption}
\usepackage{multirow}

\usepackage[subrefformat=simple, labelformat=simple]{subcaption}
\usepackage{floatrow}

\usepackage[top=2.0cm, bottom=3cm, left=2.5cm, right=2.5cm]{geometry}

\newenvironment*{mytitle}{\begin{LARGE}\bf}{\end{LARGE}\\[1.5ex]}%
\newenvironment*{myabstract}{\begin{Large}\bf}{\end{Large}\\[2.5ex]}%
\makeatletter
{}

\makeatother

\begin{document}

\begin{mytitle} Towards Monocular Digital Elevation Model (DEM) Estimation by Convolutional Neural Networks - Application on Synthetic Aperture Radar Images \end{mytitle}

Gabriele Costante, \,  University of Perugia, Dep. of Engineering, gabriele.costante@unipg.it, Italy\\
Thomas A. Ciarfuglia \,  University of Perugia, Dep. of Engineering, thomas.ciarfuglia@unipg.it, Italy\\
Filippo Biondi, \, University of L'Aquila,  Dep. of Engineering biopippoo@gmail.com, Italy\\

\input{settings/def_macros.tex}


\vspace{5ex} 

\begin{myabstract} Abstract \end{myabstract}
Synthetic aperture radar (SAR) interferometry (InSAR) is performed using repeat-pass geometry. InSAR technique is used to estimate the topographic reconstruction of the earth surface. The main problem of the range-Doppler focusing technique is the nature of the two-dimensional SAR result, affected by the layover indetermination. In order to resolve this problem, a minimum of two sensor acquisitions, separated by a baseline and extended in the cross-slant-range, are needed. 
However, given its multi-temporal nature, these techniques are vulnerable to atmosphere and Earth environment parameters variation in addition to physical platform instabilities. Furthermore, either two radars are needed or an interferometric cycle is required (that spans from days to weeks), which makes real time DEM estimation impossible. In this work, the authors propose a novel experimental alternative to the InSAR method that uses single-pass acquisitions, using a data driven approach implemented by Deep Neural Networks. We propose a fully Convolutional Neural Network (CNN) Encoder-Decoder architecture, training it on radar images in order to estimate DEMs from single pass image acquisitions. Our results on a set of Sentinel images show that this method is able to learn to some extent the statistical properties of the DEM. The results of this exploratory analysis are encouraging and open the way to the solution of single-pass DEM estimation problem with data driven approaches. 

\vspace{4ex}	

\begin{multicols}{2}

\section{Introduction} \label{sec:introduction}

Interferometric Synthetic Aperture Radar (InSAR) allows topographic reconstruction of a physical environment. The technique is performed designing a spatial single-baseline SAR geometry \cite{madsen1992automated},  \cite{bovenga2014multi},\cite{zebker1986topographic} where the result is a digital elevation model (DEM). However, to solve the phase indetermination with a good altitude accuracy, a minimum of two pass are needed, and this usually implies that we need to wait days, or months between the first and the second pass. 
We propose a method for estimating the topographic reconstruction with machine learning, implemented by Convolutional Neural Networks (CNNs) in order to estimate a DEM using only one single-look-complex (SLC) SAR image. Before getting inside the description of the novel signal processing technique, a brief analysis of the InSAR history is given. 
It is necessary to go back in time, until 1980, where Walker \textit{et al.} in \cite{walker1980range} admits the feasibility of fine Doppler frequency resolution existing for the range-Doppler SAR image. In this context a high energy scattering point target may move through several range-Doppler resolution cells, producing a smeared trace. 

SAR data are represented with a three-dimensional Fourier transform of the object reflectivity density. A full three-dimensional environment reconstruction is processable by an inverse Fourier transform. Munson \textit{et al.} \cite{munson1983tomographic} show that spotlight SAR, interpreted as a tomographic reconstruction problem, synthesizes high resolution terrain maps observed along multiple observation angles. 
\begin{figure}[H]
 	\centering
 	\includegraphics[width=0.9\columnwidth]{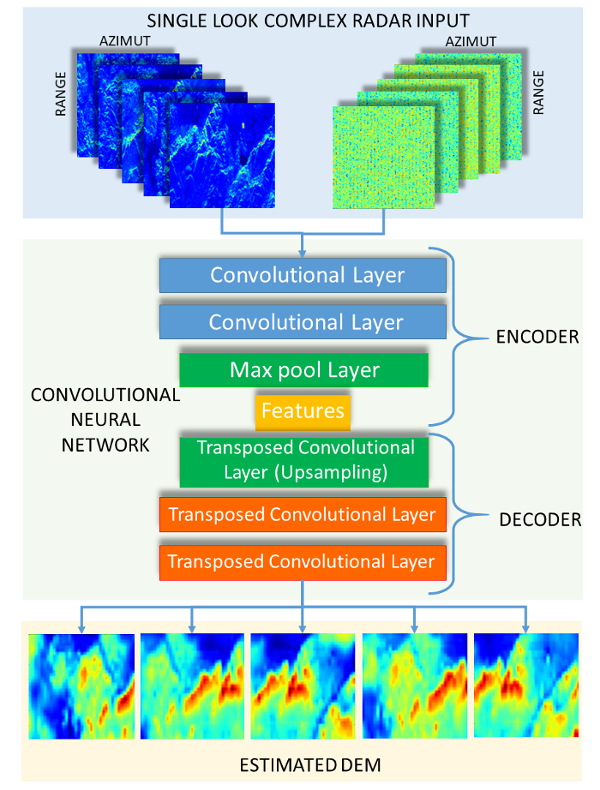}
 	\caption{\small Overview of the proposed DEM estimation approach. The proposed approach does not need multiple radar acquisitions. Instead, it processes a single look complex radar image to estimate the associated elevation model. In order to achieve this, we exploit the convolutional neural network paradigm. In particular, the encoder section extracts highly informative local structures (\ie, features) from the input radar image. Afterwards, the decoder section decodes the features and predicts the DEM image.}
 	\label{fig:overview}
 \end{figure}
Jakowats \textit{et al.} \cite{jakowatz1995new} extend the work of Munson \textit{et al.} to a new three-dimensional formulation, making the simplifying assumption that the SAR range-Doppler image is two-dimensional. Unfortunately, this assumption implies the generation of the \textit{layover effect} and, in order to explore target detection in the cross-slant range, multiple observations have to be performed. 
In \cite{biondi2016mca} the author gave a theoretical explanation of the frequency diversity in SAR-Tomography. A very good introduction to InSAR is given in \cite{margarit2007single}. The work gives detailed information for combining complex SAR images recorded by antennas positioned at different locations.
Recent years saw a refinement of the InSAR technique, trying to remove the need of using multiple satellite passes. InSAR can also be applied using two sensors mounted on the same platform. This configuration is called single-pass interferometry \cite{margarit2007single}. However, to obtain a digital elevation model with useful accuracy a minimum baseline is needed. Application of InSAR from spaceborne radar prospective is also given in \cite{moreira1995x}. In Colasanti \textit{et al.} \cite{colesanti2003generation} authors performed a precious study regarding ERS-ENVISAT interferometry despite of their carrier frequency having a shift of 31MHz. In \cite{duque2015absolute} the authors gave demonstration in estimating absolute height using a single staring spotlight SAR image using the information of different azimuth defocusing levels generated by scatterers positioned at different heights. 
The problem of this technique seems being excessively anchored to the nature of the staring-spotlight acquisition which gives a reduced range-azimuth swath of observation and precious absolute height estimation is possible only for few azimuth intra-chromatic high coherency scatterers.
However, all the aforementioned methods require complex models and computations to take into account all the atmosphere, sensor and environment conditions. Up to the authors knowledge, the possibility of computing DEM estimates with a standard SAR sensor and with a single-pass acquisition has not be tackled before. In this work, we propose the use of a different paradigm to solve this problem. Since a lot of SAR images has been collected in the past, we adopt a data driven approach. The work has been inspired by recent work on Monocular Depth Estimation performed in the Robotics and Computer Vision communities \cite{mancini2016iros}, \cite{mancini2017ral}, \cite{saxena2006learning}, \cite{saxena2009make3d}, \cite{eigen2014depth}, \cite{eigen2015predicting}, \cite{roymonocular}, \cite{liu2015learning}. Usually, in the Robotics context, depth estimation from standard camera sensors is done by triangulation of information collected through stereo-rigs, or using multiple passes of the same sensor. Recently, Convolutional Neural Networks (CNNs) models have been proposed to perform a reconstruction of a depth map from a single image acquisition. The problem of learning depth from image appearance has similarities with the task of learning DEMs from radar images. 
In this work, we propose to use the same reasoning, learning the conditional distribution of digital elevation maps from single-pass interferometric imaging. We show that the proposed model is able to learn to some extent the spatial relationships from the input data, even with a moderate amount of data. This preliminary study already shows promising results for future developments. 

\section{Methodology} \label{sec:methodology}

In order to perform DEM estimation from single-pass SAR acquisition, we need to infer the structure of the observed Earth portion by only using a single radar image. 
We achieve this by devising a deep neural network architecture that learns to predict the DEM by extracting structures
and high-level information from the input radar image. 
The key intuition behind this strategy lies in the exploitation of local image structures to infer the DEM value at a certain location (\ie, image pixel).
By using multiple stages of convolutional filters, we are able to extract high-level structures (\ie, features) at different scales. These features are then used by
the model to resolve ambiguities and estimate the DEM.

In the remainder of this section, we firstly describe more formally the principles behind our approach. Afterwards, we provide details about the proposed convolutional neural network architecture.

\subsection{Estimation Problem Formulation}

We want to model a function $\mbf{f}$ that, given a single radar image $\mathcal{I} \in \mathbb{C}^{n \times m}$ represented in the complex range-azimuth domain, is able to estimate the relative DEM, filtering out radar noise and resolving the layover indetermination. The output of the model is the DEM image $\mathcal{D} \in \mathbb{R}^{n \times m}$, where each entry contain the elevation value at that location.
In order to evaluate the contribution of the complex components of the radar image, we give the model as input the absolute value $\mathcal{I_{\rho}} \in \mathbb{R}^{n \times m} = abs(\mathcal{I})$ and the phase $\mathcal{I_{\phi}} \in \mathbb{R}^{n \times m} = phase(\mathcal{I)}$ of the complex image $\mathcal{I}$.
Thus, our function is defined as $\mbf{f}:\mathcal{I_{\rho}}, \mathcal{I_{\phi}} \rightarrow \mathcal{D}$.

\begin{figure*}[t!]
 	\centering
 	\includegraphics[width=\linewidth]{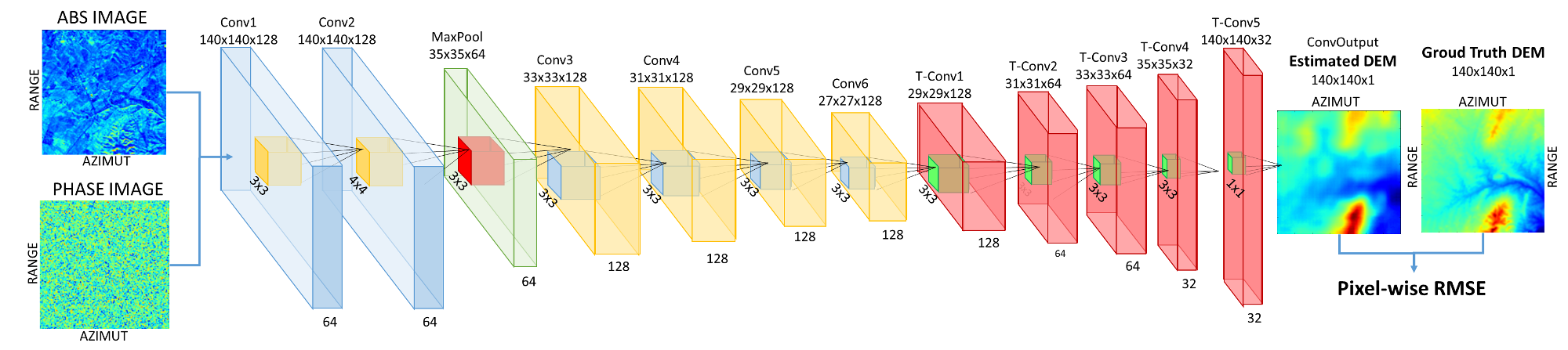}
 	\caption{\small Overview of the proposed Convolutional Neural Network DEM estimator. The encoder section processes the input radar image (\ie, the absolute and the phase images) and extracts features at different scales to detect informative local structures. The decoder sections decodes the features to estimate the associated DEM image.}
 	\label{fig:overview_cnn}
 \end{figure*}

\begin{table*}[]
\caption{\small Details of the network architecture. The network is composed by two sections, namely the encoder and the decoder section. The encoder section has six convolutional filters and a max pooling layer to extract features at different scale levels. The decoder section decodes the features by using transposed convolutions to estimate the DEM image. Padding \textit{same} is used when we need to preserve the dimensions of a layer input. Conversely, padding \textit{valid} indicates that the convolution operation processes only valid patch of the input (\ie, the output dimension is slightly smaller due to border effects).}
\label{tab:network-param}

\centering

\resizebox{0.7\textwidth}{!}{

\begin{tabular}{|c||c|c|c|c|c|c|}

\cline{1-7}
{} & {Layer name} & {Kernel size} & {Stride} & {Padding} & {output size} & {activation}  \\
\hline
Input & - & - & - & - & $(140, 140, 2)$ & \\
\hline
\multirow{3}{*}{Encoder Section} & Conv1 & $3 \times 3 \times 64 $ & $1 \times 1$ & same & $(140, 140, 64)$ & ReLU\\
\cline{2-7}
{} & Conv2 & $5 \times 5 \times 64 $ & $1 \times 1$ & same & $(140, 140, 64)$ & ReLU\\
\cline{2-7}
{} & MaxPool & $4 \times 4 \times 64 $ & $4 \times 4$ & - & $(35, 35, 64)$ & - \\
\cline{2-7}
{} & Conv3 & $3 \times 3 \times 128 $ & $1 \times 1$ & valid & $(33, 33, 128)$ & ReLU\\
\cline{2-7}
{} & Conv4 & $3 \times 3 \times 128 $ & $1 \times 1$ & valid & $(31, 31, 128)$ & ReLU\\
\cline{2-7}
{} & Conv5 & $3 \times 3 \times 128 $ & $1 \times 1$ & valid & $(29, 29, 128)$ & ReLU\\
\cline{2-7}
{} & Conv6 & $3 \times 3 \times 128 $ & $1 \times 1$ & valid & $(27, 27, 128)$ & Linear\\
\hline
\multirow{3}{*}{Decoder Section} & T-Conv1 & $3 \times 3 \times 128 $ & $1 \times 1$ & valid & $(29, 29, 128)$ & PReLU\\
\cline{2-7}
{} & T-Conv2 & $3 \times 3 \times 64 $ & $1 \times 1$ & valid & $(31, 31, 64)$ & PReLU\\
\cline{2-7}
{} & T-Conv3 & $3 \times 3 \times 64 $ & $1 \times 1$ & valid & $(33, 33, 64)$ & PReLU\\
\cline{2-7}
{} & T-Conv3 & $3 \times 3 \times 32 $ & $1 \times 1$ & valid & $(35, 35, 32)$ & PReLU\\
\cline{2-7}
{} & T-Conv4 & $3 \times 3 \times 32 $ & $4 \times 4$ & valid & $(140, 140, 32)$ & PReLU\\
\cline{2-7}
{} & ConvOutput & $3 \times 3 \times 1 $ & $1 \times 1$ & same & $(140, 140, 1)$ & Linear\\
\hline
\end{tabular}

} 

\end{table*}

For the network structure we exploit the \textit{encoder-decoder} paradigm, similar to \cite{badrinarayanan2015segnet,fischer2015flownet, mancini2016iros, mancini2017ral}. This kind of architecture is composed by two main blocks, each one composed by a number of convolutional layers, as shown in Figure \ref{fig:overview}. The encoder part computes the spatial features and at the same time reduces the image representation size layer after layer, in order to find an encoded representation of the image; the decoder part takes this encoded representation ad decompresses it, with upsamplings and convolutions, to finally reconstruct the original image. The loss that is minimized is the DEM reconstruction error, that is propagated through the decoder and encoder layers. In this way, the network is able to learn a lower dimensional representation (an embedding) of the input radar images, removing noise and increasing the generalization properties for further processing. 
We propose to use a variation of Encoder-Decoder architecture where the input and output are not the same. In our case the inputs are radar images and the outputs are the DEM reprojected in the radar coordinates (slant-range versus azimuth).

The architecture we propose is a fully convolutional deep network, that is able to handle generic inputs. Furthermore, fully convolutional architectures preserve the spatial information both in the encoder and the decoder sections, which is crucial to fully exploit local structure information.

The encoder section is composed by a series of convolutional layers, which sequentially apply learned filters on their input to compute the features. 
%
%

To extract higher level features, the input is downsampled multiple times. To scale inputs, we use max pooling.

The decoder section is composed by a stack of transposed convolutional layers that learn to reconstruct the pixel-wise predictions of the DEM image from the features computed in the encoder section. Differently from the encoder section, instead of using unpooling layers to reverse pooling operations, we take advantage from the transposed convolutional layers to learn an effective upsampling strategy. 

The network is detailed in Table \ref{tab:network-param} and shown in Figure \ref{fig:overview_cnn}. 
All the convolutional layers in the encoder section have rectified-linear activation functions (ReLU), except for the last one (Conv6) that has a linear activation function.

The decoder section has five $3\times 3$  Transposed Convolutional (T-Conv) layers to decode the feature extracted by the first section of the network. The last T-Conv layer performs an upsampling by striding the convolutional operations by a factor of 4. All the T-Conv have probabilistic rectified-linear unit (PReLU) to allow for negative activations during the decoding phase.
Finally, a single channel $3\times 3$  convolutional layer with a linear activation outputs the predicted DEM image. 

All the convolutional filters are regularized with L2 penalty to prevent overfitting.

The objective that is minimized during the learning phase is the pixel-wise linear root mean squared error (RMSE) between the estimated and the GT-DEM images: 

\begin{equation} \label{eq:log_rmse}
	\sqrt{\frac{1}{T}\sum_{i=0}^{T}||d^{gt}_{i}-\hat{d}_{i}||^{2}}  
\end{equation} 
where $T$ is the number of pixel of the DEM image $d^{gt}_{i} \in \mathcal{D}_{gt}$ and $\hat{d}_{i} \in \mathcal{\hat{D}}$.

\section{Experiments} \label{sec:experiments}

In this section, we describe the experiments we run to validate our proposed CNN-based DEM estimation approach. In the following, we first describe the experimental setup, providing details about datasets used, preprocessing procedures and details about CNN training.
Afterwards, we discuss the results and draw conclusions.

\subsection{Datasets} \label{sec:dataset}
\begin{figure}[H]
\centering
  \ffigbox{}
  {
    \CommonHeightRow
    {
      \begin{subfloatrow}[2]
        \ffigbox[\FBwidth]
        {\includegraphics[height=\CommonHeight]{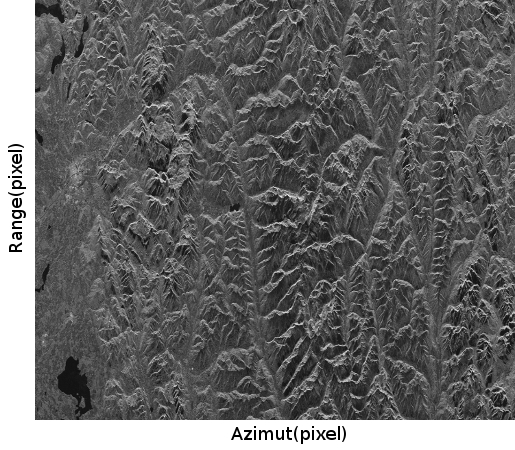}}
        {\caption{\scriptsize SLC - Alps}\vspace{-1em}\label{fig:dataset_alps_slc}}
        \ffigbox[\FBwidth]
        {\includegraphics[height=\CommonHeight]{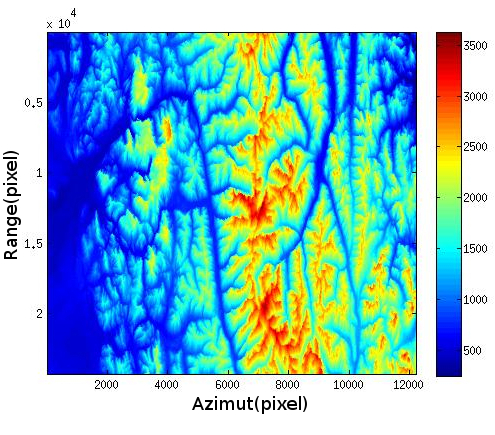}}
        {\caption{\scriptsize GT DEM - Alps}\vspace{-1em}\label{fig:dataset_alps_dem}}
        
      \end{subfloatrow}
    }   
    \CommonHeightRow
    {
     \begin{subfloatrow}[2]
        \ffigbox[\FBwidth]
        {\includegraphics[height=\CommonHeight]{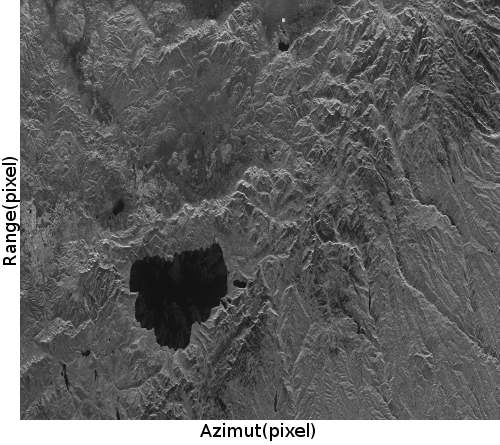}}
        {\caption{\scriptsize SLC - California}\vspace{-1em}\label{fig:dataset_sf_slc}}
        \ffigbox[\FBwidth]
        {\includegraphics[height=\CommonHeight]{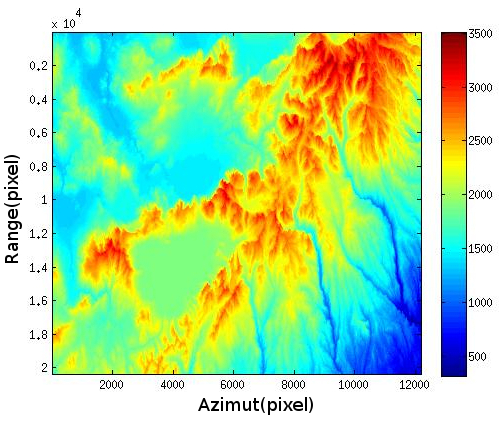}}
        {\caption{\scriptsize GT DEM - California}\vspace{-1em}\label{fig:dataset_sf_dem}}
        \end{subfloatrow}
    }   
    \CommonHeightRow
    {
     \begin{subfloatrow}[2]
        \ffigbox[\FBwidth]
        {\includegraphics[height=\CommonHeight]{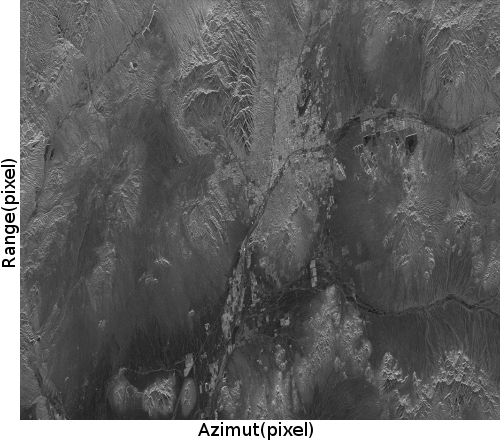}}
        {\caption{\scriptsize SLC - Tucson}\vspace{-1em}\label{fig:dataset_tucson_slc}}
        \ffigbox[\FBwidth]
        {\includegraphics[height=\CommonHeight]{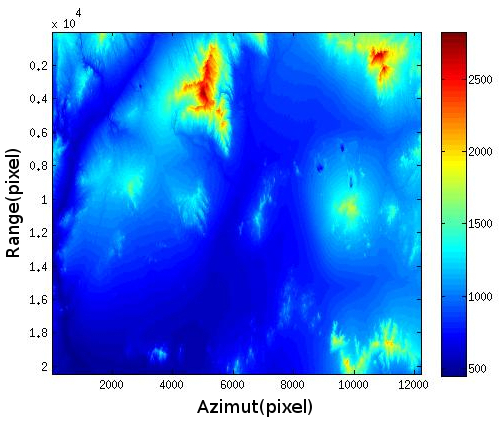}}
        {\caption{\scriptsize GT DEM - Tucson}\vspace{-1em}\label{fig:dataset_tucson_dem}}
        \end{subfloatrow}
    }   

      \caption{\small The datasets used for validating the proposed approach. The first row refers to the Alps dataset, the second to the California dataset and the third to the Tucson dataset. The first column depicts the SLC images, while the second one shows the associated GT DEM.}
        \label{fig:datasets}
      }
\end{figure}

We test out approach in three different datasets, namely the Alps, the California and the Tucson datasets. The SLC image and the associated GT DEM are depicted in Figure \ref{fig:dataset_alps_slc}-\ref{fig:dataset_alps_dem}, \ref{fig:dataset_sf_slc}-\ref{fig:dataset_sf_dem} and \ref{fig:dataset_tucson_slc}-\ref{fig:dataset_tucson_dem}, respectively. 
These datasets are taken from the Sentinel European Space Agency satellite mission. In particular, we use three different acquisitions observing the Alps (Italy), California (USA) and the city of Tucson (USA). The datasets are Single Look Complex (SLC) and vertical-vertical (VV) polarized.
Each acquisition is composed by an SLC image with the associated DEM (GT-DEM) computed with standard InSAR techniques. The SLC images are provided as a big complex matrix (typically 12000x20000 entries), while the GT-DEM is a real valued matrix with the same size of its corresponding SLC image.

To learn the CNN model, we generate the training and test samples by sliding a 4000x4000 window on the SLC/GT-DEM pair. The window has a step of 100 pixels with respect to both row and column directions. The size of the window is chosen so that each sample contains enough local structure information to allow the CNN to properly estimate the DEM image. Each sample is downsampled to 140x140 pixels to make the learning task tractable. 
Depending on the size of the input matrices, we generate up to 22000 samples for each dataset (the exact sample number is discussed in the following sections).
The train-test split is generated by randomly selecting the 65\% of samples for training and the 35\% for testing.

\subsection{Training details} \label{sec:training_details}

The CNN network is trained by using the Adam Optimizer, setting the learning rate $\alpha = 0.001$, the exponential decay rates for the moment estimates $\beta_1=0.9$ and $\beta_2=0.999$, and $\epsilon = 10^{-08}$. All the L2 regularizer values of the convolutional layer are set to $0.01$.
The batch size is set to 128 for all the experiments and the training set is randomly shuffled at the end of each epoch.
Each model is trained for 500 epochs, which takes approximately four hours with a desktop workstation equipped with a Titan Xp GPU. 
Once the model is learnt, the predictions run very fast at test time: the computation of the DEM image associated to a 4000x4000 SLC subwindow takes $0.022$ms, \ie it runs at approximately $450$ Hz.

\subsection{Discussion} \label{sec:discussion}

\begin{figure*}[ht]
\centering
  \ffigbox{}
  {
    \CommonHeightRow
    {
      \begin{subfloatrow}[4]
        \ffigbox[\FBwidth]
        {\includegraphics[height=\CommonHeight]{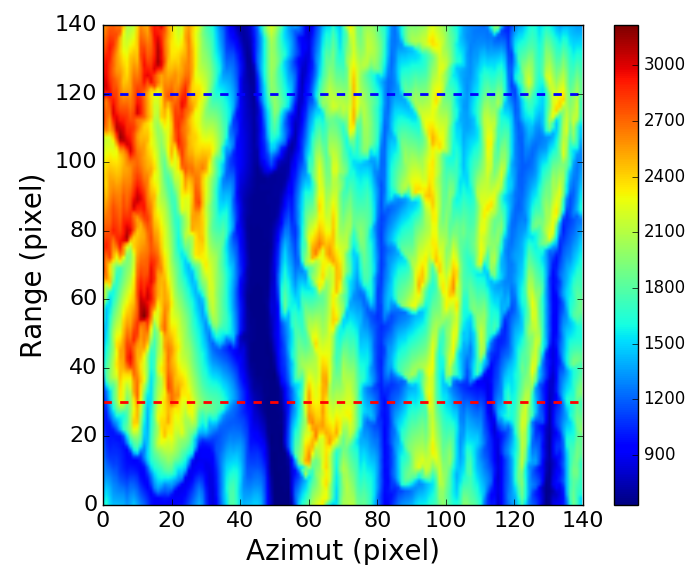}}
        {\caption{\scriptsize GT DEM - Alps test}\vspace{-1em}\label{fig:gt_2250_alps}}
        \ffigbox[\FBwidth]
        {\includegraphics[height=\CommonHeight]{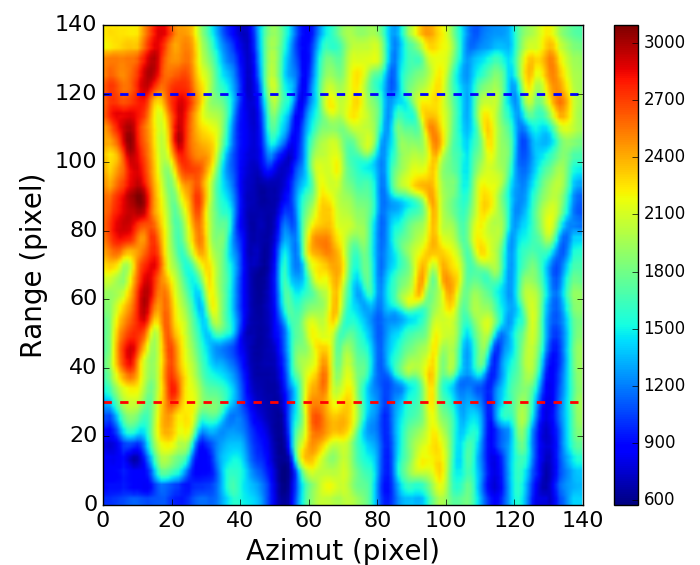}}
        {\caption{\scriptsize Est - DEM - Alps test}\vspace{-1em}\label{fig:pred_2250_alps}}
       \ffigbox[\FBwidth]
        {\includegraphics[height=\CommonHeight]{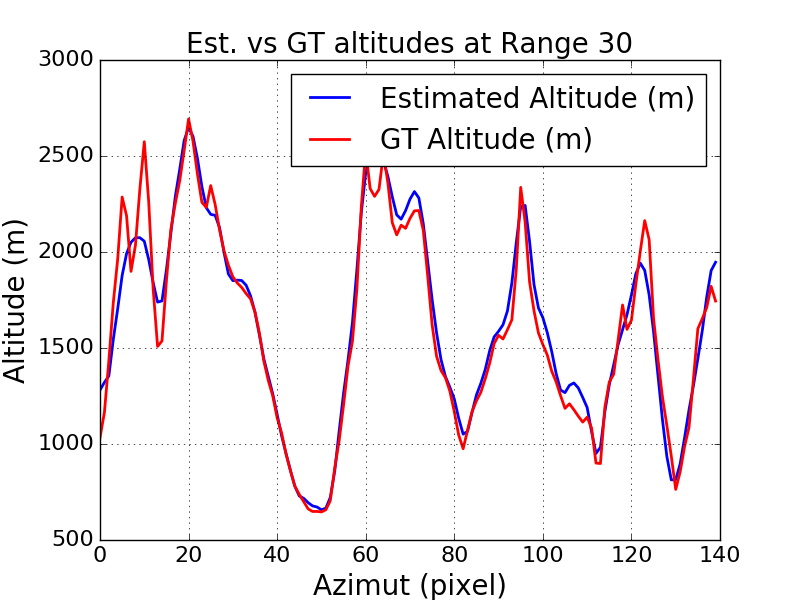}}
        {\caption{\scriptsize Est. vs GT altitudes \\Range - 30 Alps test}\vspace{-1em}\label{fig:az_30_2250_alps}}
        \ffigbox[\FBwidth]
        {\includegraphics[height=\CommonHeight]{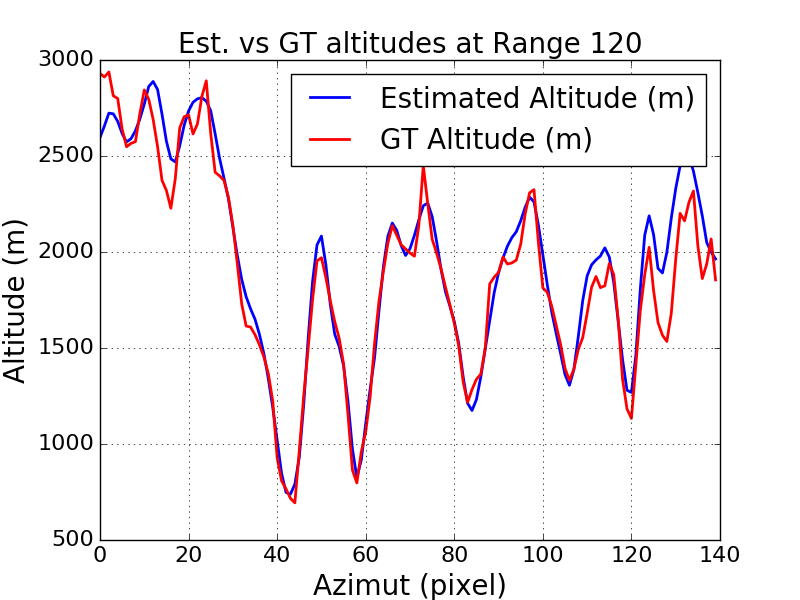}}
        {\caption{\scriptsize Est. vs GT altitudes \\Range 120 - Alps test}\vspace{-1em}\label{fig:az_120_2250_alps}}
        
      \end{subfloatrow}
    }   
    \CommonHeightRow
    {
     \begin{subfloatrow}[4]
        \ffigbox[\FBwidth]
        {\includegraphics[height=\CommonHeight]{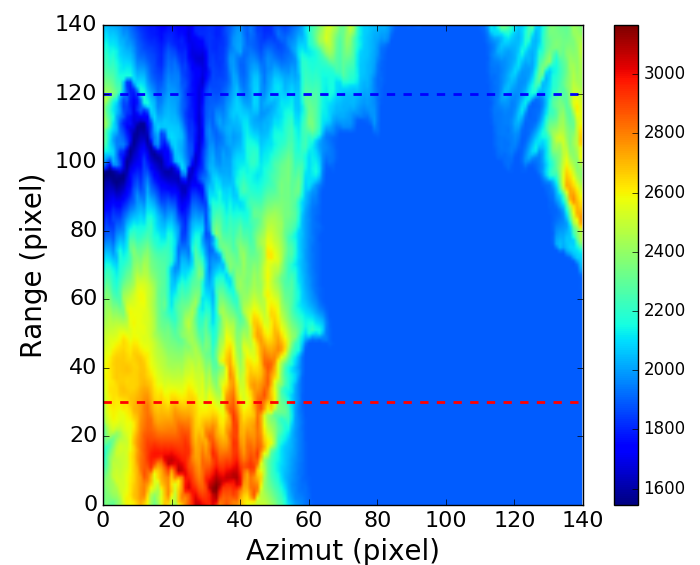}}
        {\caption{\scriptsize GT DEM - California test}\vspace{-1em}\label{fig:gt_3700_sf}}
        \ffigbox[\FBwidth]
        {\includegraphics[height=\CommonHeight]{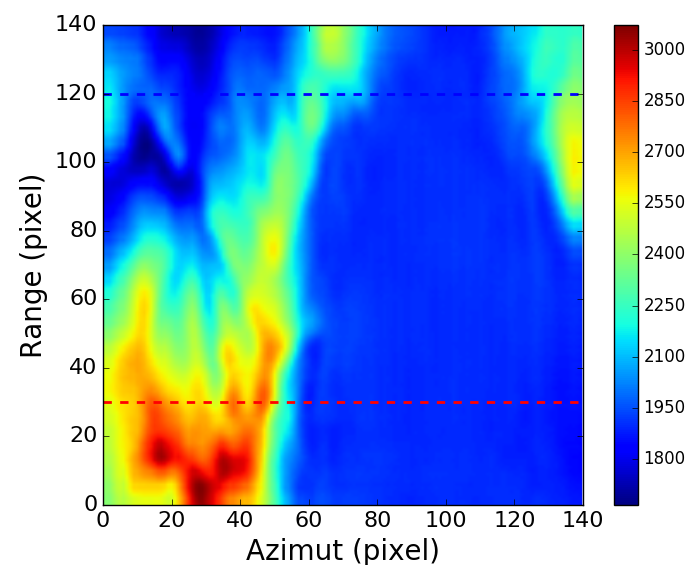}}
        {\caption{\scriptsize Est - DEM - California test}\vspace{-1em}\label{fig:pred_3700_sf}}
        \ffigbox[\FBwidth]
        {\includegraphics[height=\CommonHeight]{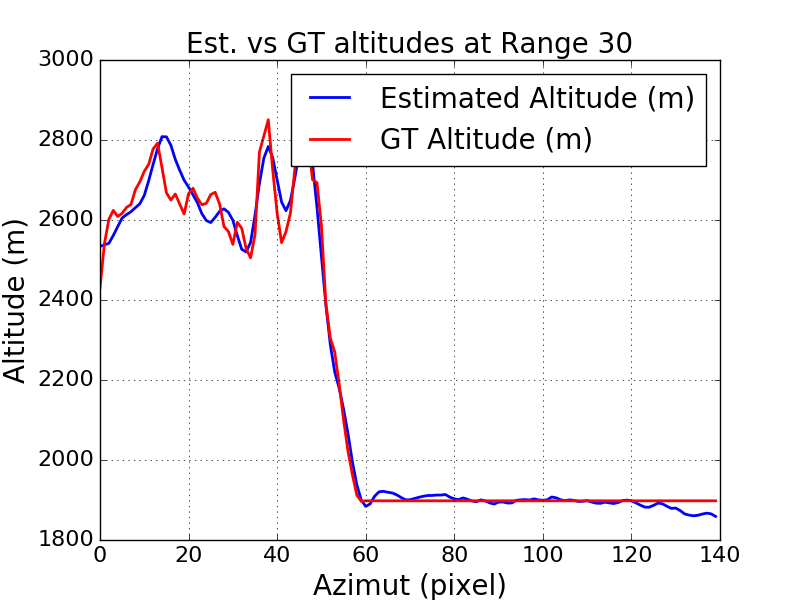}}
        {\caption{\scriptsize Est. vs GT altitudes \\Range 30 - California test}\vspace{-1em}\label{fig:az_30_3700_sf}}
        \ffigbox[\FBwidth]
        {\includegraphics[height=\CommonHeight]{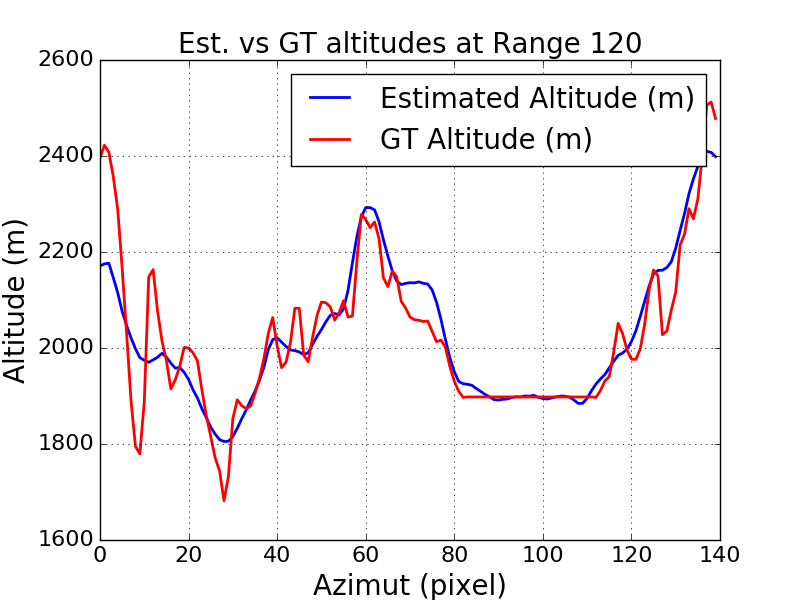}}
        {\caption{\scriptsize Est. vs GT altitudes \\Range 120 - California test}\vspace{-1em}\label{fig:az_120_3700_sf}}
        \end{subfloatrow}
    }   
    \CommonHeightRow
    {
     \begin{subfloatrow}[4]
        \ffigbox[\FBwidth]
        {\includegraphics[height=\CommonHeight]{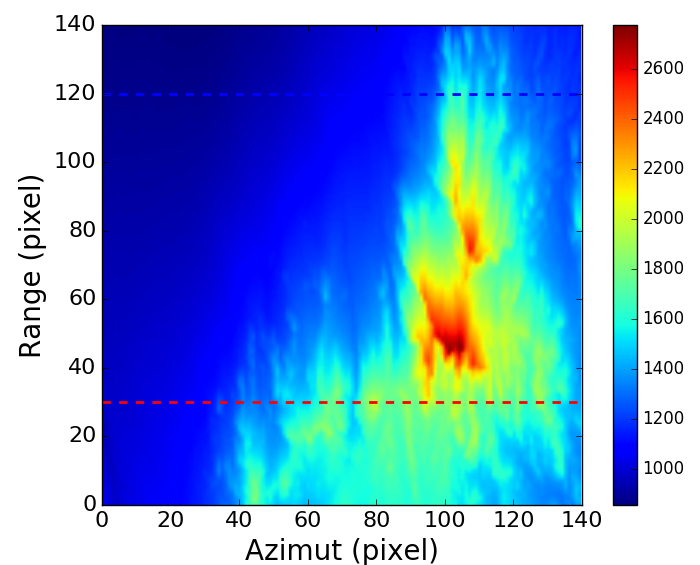}}
        {\caption{\scriptsize GT DEM - Tucson test}\vspace{-1em}\label{fig:gt_25_tucson}}
        \ffigbox[\FBwidth]
        {\includegraphics[height=\CommonHeight]{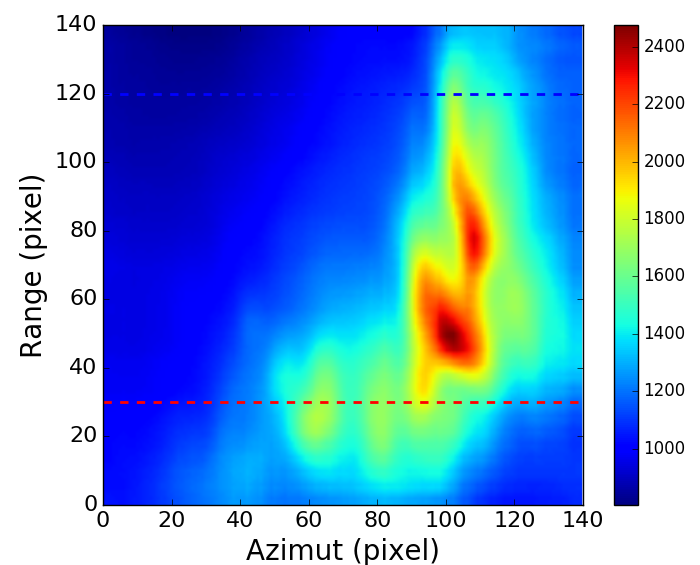}}
        {\caption{\scriptsize Est - DEM - Tucson test}\vspace{-1em}\label{fig:pred_25_tucson}}
        \ffigbox[\FBwidth]
        {\includegraphics[height=\CommonHeight]{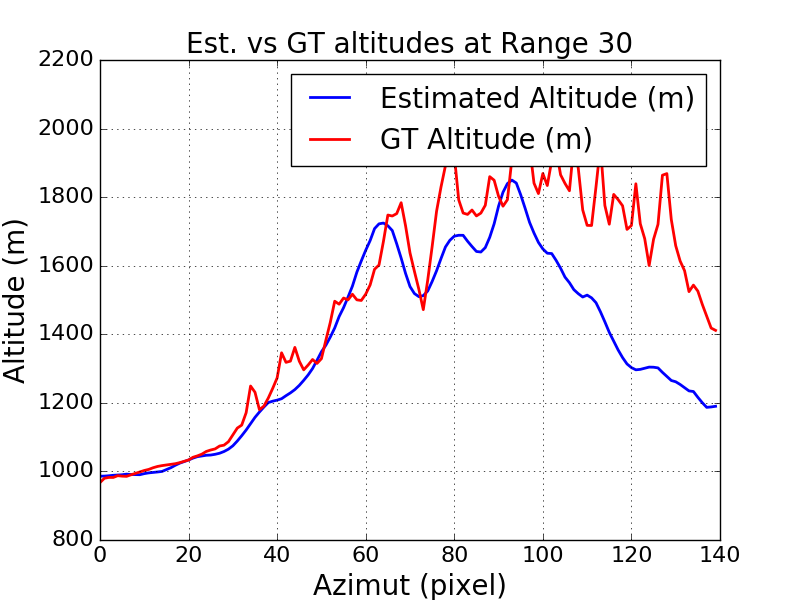}}
        {\caption{\scriptsize Est. vs GT altitudes \\Range 30 - Tucson test}\vspace{-1em}\label{fig:az_30_25_tucson}}
        \ffigbox[\FBwidth]
        {\includegraphics[height=\CommonHeight]{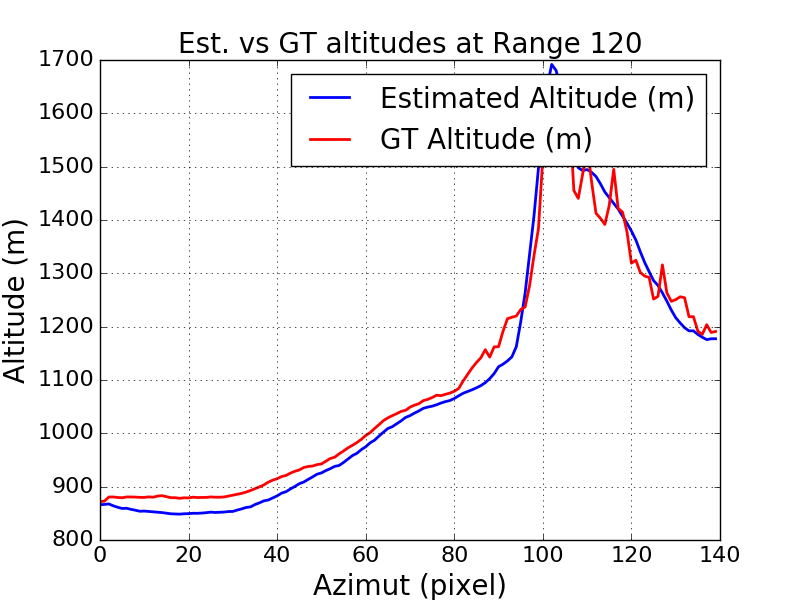}}
        {\caption{\scriptsize Est. vs GT altitudes \\Range 120 - Tucson test}\vspace{-1em}\label{fig:az_120_25_tucson}}
        \end{subfloatrow}
    }   

      \caption{\small Comparison between the estimated and the ground truth DEM images. The first row refers to the experiment on the Alps dataset, while the second and the third show the results for the California and the Tucson tests, respectively. The first column depicts the GT DEM of a sample image from the three datasets, while the second column shows the relative estimated DEM. The third and the fourth columns compare the estimated altitude profiles with the ground truth ones at fixed range values.}
        \label{fig:fixed_azimut_exp}
      }
\end{figure*}

Figure \ref{fig:fixed_azimut_exp} shows examples of the real DEMs and the estimated ones for each datasets. In addition, the elevation profiles are plotted for two sample range (in pixel), in order to better show the estimation properties of the network. Alps dataset is composed by $11031$ train images and $5818$ test images, and the average RMSE on all test images is $105.28$m. California is composed by $8693$ train images and  $4755$ test images. The average RMSE in this case is $74.46$m. Finally, the Tucson dataset has $8846$ train and $4849$ test images. In this test, the average RMSE error is $43.45$m.

\begin{figure*}[ht]
   \centering
  \ffigbox{}
  {
    \CommonHeightRow
    {
      \begin{subfloatrow}[3]
        \ffigbox[\FBwidth]
        {\includegraphics[height=\CommonHeight]{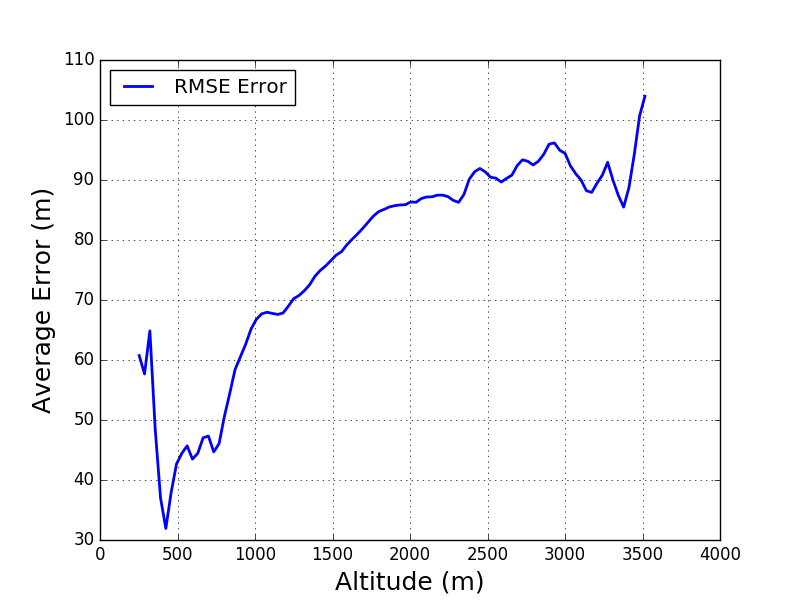}}
        {\caption{\scriptsize Avg Estimation error on the ALPS data}\vspace{-1em}\label{fig:alt_error_alps}}
        \ffigbox[\FBwidth]
        {\includegraphics[height=\CommonHeight]{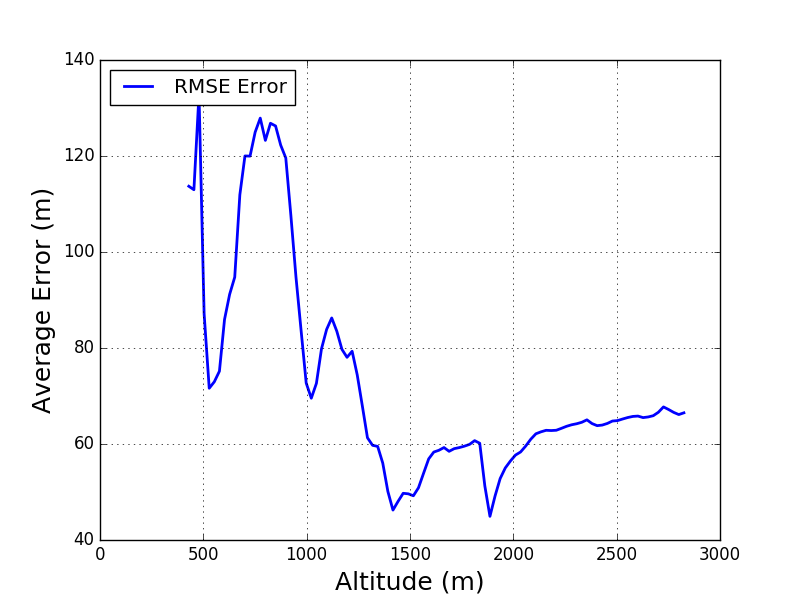}}
 	    {\caption{\scriptsize Avg Estimation error on the California data}\vspace{-1em}\label{fig:alt_error_sf}}
        
        \ffigbox[\FBwidth]
        {\includegraphics[height=\CommonHeight]{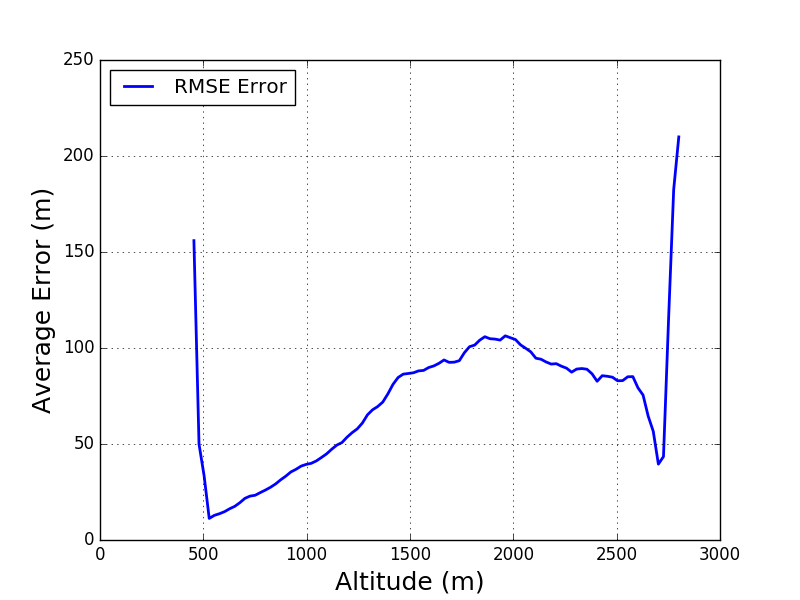}}
 	    {\caption{\scriptsize Avg Estimation error on the Tucson data}\vspace{-1em}\label{fig:alt_error_tucson}}
      \end{subfloatrow}
    }   

      \caption{\small Average estimation error computed on quantized elevations (100 bins) on the ALPS \ref{fig:alt_error_alps}, the California \ref{fig:alt_error_sf} and the Tucson \ref{fig:alt_error_tucson} data.}
        \label{fig:exps_alps2}
      }
\end{figure*}

It is possible to see qualitatively that the Network learned the altitude statistics, giving a result that closely resembles the ground truth. The main difference is a smoothing effect that the network estimate has in comparison with the original. This is more evident if we compare the GT and predicted profiles for fixed range values (in pixel with respect to the image coordinates) of 30 and 120 pixels, respectively. This is shown in Figures \ref{fig:az_30_2250_alps}-\ref{fig:az_120_2250_alps}, \ref{fig:az_30_3700_sf}-\ref{fig:az_120_3700_sf} and \ref{fig:az_30_25_tucson}-\ref{fig:az_120_25_tucson} for the Alps, the California and the Tucson dataset, respectively.  From the profiles it is even more apparent that the network is able to output a digital elevation model for the input images that closely resembles the original. The general trends of the GT DEM are closely followed and the main differences between the GT and prediction are due to the smoothing effect on crest ripples, since that for the network these ripples in the ground truth are like a high frequency signal (noise) superimposed to the general elevation model. 
To better quantify the performances of the network, we quantized the range of elevations in the datasets and computed the average error for each bin, in order to analyse the error distribution given the GT elevation. The resulting plot is shown in Figures \ref{fig:alt_error_alps}, \ref{fig:alt_error_sf} and \ref{fig:alt_error_tucson} for the Alps, the California and the Tucson datasets, respectively. 
The three plots, together with the ones in Figure \ref{fig:fixed_azimut_exp} and in consideration of the average RMSE on the test sets show that the estimation network performances degrades when the terrain is mountainous, while are close to the real DEM for slow varying terrain features. This is expected, since the altitude information is not really included in a single pass radar image, so the Network has to extract it from context level information. We hypothesize that, increasing the amount of data given to the network, is possible to further reduce the errors on the high frequency ripples. Furthermore, devising more complex architectures should also help to better model the variabilities of high crests.  

\section{Conclusions}
In this paper, we have proposed an novel method able to estimate DEMs using single SAR images instead interferometric couples. The proposed method uses a data driven approach, implemented through an \textit{Encoder-Decoder} CNNs architecture, and is able to potentially solve the layover indetermination present on the single SLC SAR image using image context information. Our results show that this  method is promising, and able to learn useful DEM estimate even with moderate training time and data. For training the CNN a set of Sentinel data has been used.


\nocite{*}
\bibliographystyle{plain}

\bibliography{DeepMonoSAR_bibliography}

\end{multicols}
\end{document}